\documentclass[aps,showpacs,amsmath,amssymb,showkeys,twocolumn,final]{revtex4-1}

\usepackage{graphicx}

\begin{document}
	
	\title{ The role of Coulomb interaction in superconducting NbTiN thin films}
	
	\author{D. Hazra$^{1}$, N. Tsavdaris$^{2}$, A. Mukhtarova$^{1}$, M. Jacquemin$^{2}$, F. Blanchet$^{1}$, R. Albert$^{1}$, S. Jebari $^{1}$, A. Grimm$^{1}$, E. Blanquet$^{2}$, F. Mercier$^{2}$, C. Chapelier$^{1}$ and M. Hofheinz$^{1}$}
	
	\affiliation{$^{1}$ Univ.\ Grenoble Alpes, CEA, INAC-Pheliqs, 38000 Grenoble, France }
	\affiliation{$^{2}$ Univ.\ Grenoble Alpes, CNRS, Grenoble INP, SIMaP,  38000 Grenoble, France}
	
	\email []  {iamdibyenduhazra@gmail.com}
	
	\date{\today}

	\begin{abstract}
		We report on the superconducting properties of Nb$_{1-x}$Ti$_x$N thin films of thickness $\sim 10$\,nm, with different Ti fraction $x$ in the range  $ 0 \leq x \leq 0.5$, deposited by high temperature chemical vapor deposition. In this parameter range, we observe that the superconducting critical temperature ($T_c$) increases with $x$. Our analysis, in accordance with both McMillan's and Finkelstein's theories, shows that disorder-enhanced Coulomb interaction decreases with $x$, leading to an increase of $T_c$.
	\end{abstract}

		
	\maketitle

Because of its high superconducting critical temperature, high-quality NbTiN has been one of the most preferred materials for many superconducting applications, such as  superconducting coating for radio frequency cavities \cite{benvenuti1997production,fabbricatore1993niobium,bosland1993nbtin}. The high superconducting energy gap ($\Delta$) makes NbTiN very suitable for THz application, such as superconductor-insulator-superconductor mixtures and bolometers \cite{kooi1998low, bumble2001fabrication,jackson2001low,jiang2010development}. NbTiN is also a preferred material for optical single photon detection \cite{miki2013high, schuck2013waveguide, tanner2010enhanced, dorenbos2008low} and  has been used as high characteristic impedance microwave resonator \cite{samkharadze2016high}--- thanks to its very high kinetic inductance. Low loss resonators can be fabricated from NbTiN \cite{barends2010minimal,barends2009noise}; this in combination with high $\Delta$, $T_c$, and $B_{c2}$ make NbTiN a potential alternative to aluminum for circuit quantum electrodynamics in high magnetic field \cite{samkharadze2016high}.

The effect of disorder on conventional s-wave superconductivity has been extensively studied in Nb$_{1-x}$Ti$_x$N and its parent compounds, NbN and TiN \cite{goldman1998,crane2007survival,baturina2007prl,baturina2007quantum,sacepe2008prl,sacepe2010natcomm,mondal2011PRL,mondal2011role,driessen2012PRL,coumou2013PRB,mondal2013SCR,kamlapure2013emergence,noat2013unconventional}. Already vast and rich physics has been unearthed in these systems, that include superconductor-insulator transition \cite{goldman1998, baturina2007prl}, observation of a pseudogap regime above $T_c$  \cite{sacepe2010natcomm, mondal2011PRL}, disorder-induced phase fluctuation \cite{mondal2011PRL}, spatially inhomogeneous superconductivity \cite{sacepe2008prl, kamlapure2013emergence, noat2013unconventional} and enhancement of pair breaking parameter \cite{driessen2012PRL}.

Despite numerous applications and fundamental investigations the following points are clearly missing: (1) A controlled growth technique to deposit high quality thin films. (2) A clear understanding of the variation of superconducting parameters with Ti fraction ($x$). (3) A systematic way to control electronic disorder and to study its effect on superconducting properties. Here, we report on the superconducting properties of high-quality Nb$_{1-x}$Ti$_x$N films where electronic disorder can be tuned by controlling $x$. The $T_c$ of our samples increases with $x$ which we attribute to the reduction of Coulomb interaction. This is consistent with both  McMillan's and Finkelstein's  equations.

To grow Nb$_{1-x}$Ti$_x$N thin films, d.c. magnetron sputtering is the most common technique \cite{bell1968superconducting,barends2008contribution, barends2009noise,makise2011characterization,hong2013terahertz,karimi2016structural, samkharadze2016high}, but atomic layer deposition (ALD) has also been explored  \cite{klug2013heteroepitaxy}. In the case of sputtering, the high sputtering rate makes the thickness control very challenging below 10\,nm; whereas, in case of ALD, the control of both composition and crystalline quality remains difficult.

Variations of superconducting parameters, especially $T_c$, with Ti fraction $x$, have been previously reported \cite{di1990niboium}. The authors observed that $T_c$ remains almost constant up to $x \sim 0.5$ and decreases for higher values. In contrast, Myoren et al. \cite{myoren2001properties} observed a monotonous decrease of $T_c$ with $x$ for three of their films with $x = 0$, 0.34 and 0.62, respectively. In both cases, the films constituted 3d systems with thicknesses above 300\,nm and were prepared by dc magnetron sputtering. Prior to these experiments, Pressal et al. \cite{pessall1968study} and Yen et al. \cite{yen1967superconducting} observed that $T_c$ varies non-monotonically with $x$; below  $x \sim 0.4$, $T_c$ increases with $x$ and decreases above. In either of these cases, no clear explanation for the observed variation of  $T_c$ with Ti fraction was provided.

To study the effect of disorder on superconducting properties, majority of the experiments have been focussed on series of films with different thicknesses, making it difficult to disentangle bulk disorder to surface scattering contributions.

Here, to overcome these issues, we report on the superconducting properties of five Nb$_{1-x}$Ti$_x$N thin films of thickness $10$\,nm grown by high temperature chemical vapor deposition (HTCVD). The detailed structural analysis by x-ray diffraction and cross sectional high resolution transmission electron microscopy reveal that the deposited films are of very high crystalline qualities. Apart from different gas flow rates, chamber conditions are kept identical between each depositions. In this way, the only parameter changing from sample to sample is the Ti fraction ($x$), which we control in the range $ 0 \leq x \leq 0.5$ for the present study. Our goal is to understand how $x$, in the range $ 0 \leq x \leq 0.5$, impacts disorder and $T_c$. Disorder will be estimated with the Ioffe-Regel parameter $k_{F} \ell$ ($k_{F}$ is the Fermi wavevector and $\ell$ is the mean free path).

Five Nb$_{1-x}$Ti$_x$N thin films have been produced by HTCVD at 1100 $^\circ$C on Epiready (0001) oriented $\mathrm {Al_{2}O_{3}}$ substrate. Deposition apparatus and thermodynamics calculation have been reported elsewhere \cite{tsavdaris2017chemical}. Deposition conditions are the same for each sample except for the ratio of chlorine species $\mathrm{NbCl}_{x}/\mathrm{TiCl}_{x}$ in the gas phase. The control of the Nb/Ti ratio in the gas phase allows the control of the titanium concentration in the layer.
All the films are $d=10 \pm 1$\,nm thick, as determined from x-ray reflectometry.

The films are pure cubic NbTiN (ICDD: 01-088-2404); no hexagonal phases were detected. The XRD (111)-$\omega$ scan rocking curve values, referring to the tilt angle along the 111 direction between grains, are low and between 190 and 350\,arcsec with no clear dependence on Ti fraction. Thus, the crystalline quality of NbTiN is not affected by the presence of Ti. However, two NbTiN in-plane variants with an in-plane twist relationship of $60^\circ$ are detected in all samples. These in-plane variants results from the stacking of material with a cubic structure (NbTiN) on a hexagonal substrate (surface of (0001) $\mathrm {Al_{2}O_{3}}$). We found that the domains with a single variant were distributed randomly and have a lateral size in the order of 150\,nm \cite{mercier2014niobium,tsavdaris2017chemical}.


\begin{figure}\centerline{\includegraphics[width=9cm,angle=0]{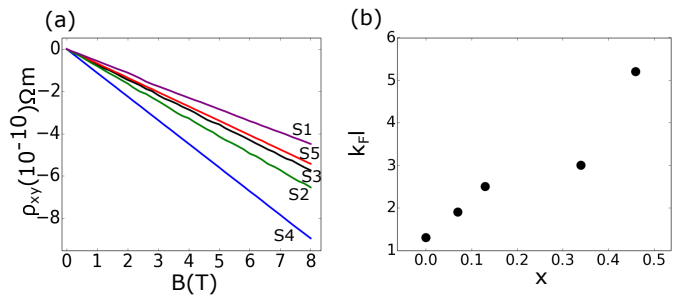}}
	\caption {(a) $\rho_{xy} $ for all five samples at 50\,K as a function of magnetic field. (b) The variation of Ioffe-Regel parameter  ($k_{F}\ell$), determined at 50\,K, as a function of Ti fraction  ($x$).}
	\label{fig:hall}
\end{figure}

Electrical transport measurements were performed in a Quantum Design Physical Property Measurement System down to 2.5\,K and up to magnetic field ($B$) 8\,T. The free electron parameters of our samples were determined from the combination of longitudinal and Hall resistivity measurement at 50\,K; lower temperatures are avoided not to be influenced by superconducting fluctuation related effects \cite{tikhonov2012fluctuation,breznay2012hall,destraz2017superconducting}. Fig.~\ref{fig:hall}a shows the variation of Hall resistivity ($\rho_{xy}$) for all five samples as a function of magnetic field. For all five samples, $\rho_{xy}$ varies linearly with magnetic field. The free electron density ($n$) is determined from the slope of the $\rho_{xy}(B)$ curve, i.e., from the Hall coefficient $R_{H} = 1/ne$, where $e$ is the charge of the electron. Knowing $n$, $k_F$ and $v_F$  (the Fermi velocity) are determined from $k_{F} = (3 \pi^{2}n)^{1/3}$  and $v_{F} = \hbar k_{F}/m $, where $m$ is the mass of the electron. The elastic scattering time ($\tau$) is estimated from Drude's formula: $\rho_{xx} = m/ne^{2}\tau$, here $\rho_{xx} = d R_S$ is the longitudinal resistivity at 50\,K. The other important free electron parameters, like, $\ell$, diffusion constant ($D$) and density of states at the Fermi level ($N_V$) are determined from $\ell = v_{F} \tau $, $D = v_{F} \ell /3$ and $N_V = mk_F/\hbar^{2}\pi^{2}$. $k_{F} \ell$ is determined from $k_{F} \ell = \frac{h}{e^2}\rho_{xx}^{-1} \left(\frac{9}{8}\pi e R_H\right)^{1/3}$.
We note that $k_{F} \ell$ depends only on experimentally measured quantities $R_{H}$ and  $\rho_{xx}$, not on effective electron mass $m$. The important free electron parameters are summarized for each sample in Table-\ref{tab1}.


\begin{table*}
	\caption{\label{tab1} An overview of some of the important parameters of our Nb$_{1-x}$Ti$_x$N thin films. The directly measured parameters and those extracted from the free electron theory are separated by the double-line.}
	\begin{tabular}{|c|c|c|c|c|c|c||c|c|c|c|c|c|c|}
		\hline
		Samples & $a$ & $x$ & RRR&  $T_{c}$&  $R_{H}$ &  $R_S$ (50\,K)  & $n$ & $\tau$ & $\ell$ & $k_F \ell$  & $D$ &  $N_{V}$     \\
		
		&($\mathrm {{\AA}}$)& & &(\,K)&($\mathrm {10^{-11} m^{3}/C}$)&($\mathrm{\Omega}$)&  ($\mathrm {10^{28}/m^{3}}$)& ($\mathrm {10^{-17}\,s}$)&($\mathrm {{\AA}}$)&(50\,K)& ($\mathrm {10^{-5} m^{2}/s}$)&  $ \mathrm {(\frac{10^{47}states}{m^{3}.J})}$  \\ \hline
		
		S1       & 4.340     &0.00&0.27& \phantom{0}7.4&\phantom{0}5.6&607&11.1&\phantom{0}5.2 &0.9& 1.3 & \phantom{0}5.1 &   1.24 \\
		S2        & 4.339     &0.07&0.37& \phantom{0}8.6&\phantom{0}8.4&502&7.4 &\phantom{0}9.6 &1.4& 1.9  & \phantom{0}7.1&  1.10\\
		S3        & 4.336     &0.14&0.55&  10.4&\phantom{0}7.9&362&7.8 &12.5 &1.9& 2.5    & \phantom{0}9.7&  1.10\\
		S4        & 4.312     &0.34&0.63& 11.4&13.8&349&4.5 &22.4 &2.8& 3.1   & 12.0 & 0.92 \\
		S5       & 4.303     &0.46&0.88&  13.1&\phantom{0}6.8&167& 9.2 &23.0 &3.7& 5.2   &19.8 & 1.17 \\
		
		\hline
	\end{tabular}
\end{table*}

In Fig.~\ref{fig:hall}b, we plot $k_{F}\ell$ as a function of $x$, showing that $k_{F}\ell$ increases monotonically with $x$. Thus, the disorder can be tuned systematically by controlling Ti fraction, making these films ideal candidates to study the effect of atomic level disorder on superconducting properties. This observation is consistent with our structural analysis \cite{tsavdaris2017chemical}, where we observed that surface morphology improved with increasing Ti fraction. This is also consistent with the fact that, both residual resistivity ratio (RRR) and $\ell$ increases with $x$ (see Table-\ref{tab1}; RRR is defined as $\rho_{xx}$ (300\,K)/$\rho_{xx}$(max)). It is well-known that RRR decreases with increasing defect density \cite{kittel2005introduction} that subsequently reduces $\ell$. At 50\,K where free electron parameters are defined, the electron-phonon interaction is small and thus the electrical resistance stems predominantly from the electron-defect scattering \cite{chand2009temperature, chand2012thesis}.

\begin{figure}\centerline{\includegraphics[width=9cm,angle=0]{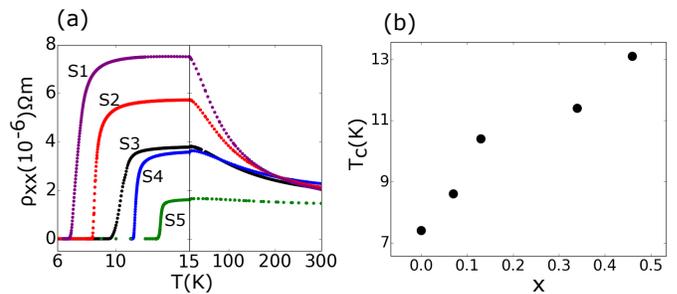}}
	\caption {(a) Temperature dependence of $\rho_{xx}$ for all five samples down to the superconducting transition temperature. (b) The variation of $T_c$ as a function of Ti fraction showing a monotonic dependence.}
	\label{fig:resistivity}
\end{figure}

 In Fig.~\ref{fig:resistivity}a,  we plot the temperature dependence of $\rho_{xx}$ at zero magnetic field. Upon cooling down from room temperature $\rho_{xx}$ increases and reaches a maximum at some intermediate temperature $T_{max}$. Below $T_{max}$, $\rho_{xx}$ starts to decrease with decreasing temperature due to the onset of superconductivity. The left panel of Fig.~\ref{fig:resistivity}a shows a magnified version near the superconducting transition. Clearly $T_c$ is systematically increasing with Ti fraction from sample S1 ($x=0$) to S5 ($x=0.5$). This is shown in Fig.~\ref{fig:resistivity}b where $T_c$ is plotted as a function of $x$. $T_c$ is defined at a temperature where $\rho_{xx}$ is half of normal resistivity defined by $\rho_{xx}$ measured at 15\,K. In Table-\ref{tab1}, we summarize $T_{c}$ of our samples.

To understand the variation of $T_c$ with $x$, we follow two different approaches: (1) McMillan's equation and (2) Finkelstein's equation.

The $T_c$ of a strongly coupled superconductor like NbTiN is governed by McMillan’s equation \cite {mcmillan68}

\begin{eqnarray}
T_{c} = \frac{\Theta_{D}}{1.14}\exp\left(-\frac{1.04(1+\lambda)}{\lambda - \mu^{*}(1+0.62\lambda)}\right).
\label{eq:Mcmillan}
\end{eqnarray}

Here, $\Theta_D$ is the Debye temperature, $\lambda$ is the effective electron-phonon coupling constant, and $\mu^{*}$ is the Coulomb pseudopotential representing electronic Coulomb repulsion. $\lambda$ is given by $\lambda = N_{V}U$, where $U$ is the attractive potential. $\Theta_D$ and $U$  depend on the phonon structure and hence lattice parameter ($a$). $\mu^{*}$, on the other hand, depends on disorder--- with increasing disorder,  $\mu^{*}$ increases \cite{anderson83}. Our five samples have different $a$, $N_V$,  and $k_{F}\ell$. Therefore, $\Theta_D$, $\lambda$, and $\mu^{*}$ are different for all five samples. Thus, it is difficult to analyze the variation of $T_c$ as a function of any of the single variables ---$a$, $N_{V}$, or $k_{F}\ell$. However, we note that the maximum change in $a$ is less than 1\,\% between our samples. Thus, the change in $\Theta_D$ from sample to sample due to change in $a$ is not enough to describe the variation of $T_c$. $N_V$, on the other hand, changes quite significantly--- the maximum variation is about 25 \%. However, we see no systematic variation of $T_c$ with $N_V$. For instance, S1 has highest $N_V$ but it also has lowest $T_c$; on the other hand, S4 has lowest $N_V$ but it has second highest $T_c$ (see Table-\ref{tab1}). In contrast, the variation of $T_c$ with $k_{F}\ell$ is more systematic. The maximum variation in $k_{F}\ell$ is about 250 \%, much more than $a$ or $N_V$. Thus, it seems that with increasing $x$, disorder of our system reduces, resulting a decrease in $\mu^{*}$. This according to Eq.\ref{eq:Mcmillan}, increases $T_c$.

\begin{figure}\centerline{\includegraphics[width=5cm,angle=0]{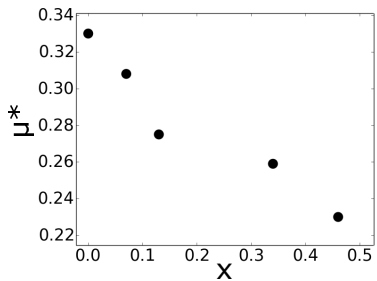}}
	\caption {Variation of the Coulomb pseudo-potential as a function of Ti fraction.}
	\label{fig:xMu}
\end{figure}

To verify this mechanism, we assume that $\lambda$ and $\Theta_{D}$ are the same for all our samples. According to Kihlstrom et al. \cite{kihlstrom1985tunneling} $\lambda = 1.46$ and $\mu^{*} = 0.33$ for NbN (S1). Substituting this in Eq.\ref{eq:Mcmillan} yields $\Theta_{D} = 183$\,K for S1. Now substituting $\Theta_{D} = 183$\,K and $\lambda = 1.46$, we determine $\mu^{*}$ from Eq.\ref{eq:Mcmillan} for the remaining samples (S2 to S5). In Fig.~\ref{fig:xMu}, we show the variation of $\mu^{*}$ as a function of $x$. The observed decrease of $\mu^{*}$ with increasing Ti fraction corresponds to decreasing Coulomb interaction, in agreement with the $k_F\ell$ parameters we extracted. 

Here, we would like to mention that $\Theta_{D} = 183$\,K is small compared to the values reported in reference \cite{chand2012thesis} and references therein which range between 250 to 350\,K. This is due to the fact that $T_c$ in reference \cite{kihlstrom1985tunneling} was 14.0\,K larger than S1. Taking $T_c = 14.0$\,K yields $\Theta_{D} = 346$\,K. Irrespective of exact value of $\Theta_{D}$, the qualitative behavior of Fig.~\ref{fig:xMu} remains the same.

Using McMillan's equation, we have argued that increase of $T_c$ with increasing $x$ can most likely be attibuted to Coulomb interaction. But, we had to assume that $\Theta_{D}$ and $\lambda$ do not change from sample to sample. We will now explore Finkelstein's formula which allows us to express $T_{c}$ in terms of experimentally measured quantities, normal sheet-resistance $R_{s}$ and $\tau$. Thus, unlike McMillan's equation, there is no free parameter in this framework.

Finkelstein's model states that with the increasing disorder the reduced scattering length reinforces the Coulomb interaction, which in turn reduces the $T_c$ from the non-disordered value according to the following equation:

\begin{eqnarray}
\frac{T_{c}}{T_{c0}} = e^{\gamma} \left(\frac{1/\gamma - \sqrt{t/2} + t/4}{1/\gamma + \sqrt{t/2} + t/4} \right)^{1/\sqrt{2t}}
\label{eq:Frinkelstein}
\end{eqnarray}

Here, $T_{c0}$ is the critical temperature for ‘non-disordered’ material, $t = R_{s}e^{2} /\pi h $, $\gamma = \ln (h/k_{B}T_{c0}\tau)$.

\begin{figure}\centerline{\includegraphics[width=9.5cm,angle=0]{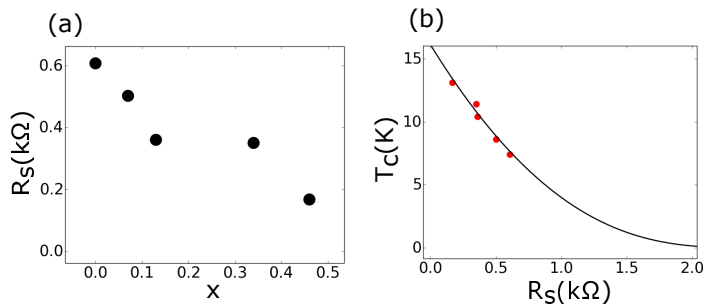}}
	\caption {(a) Variation of sheet resistance as a function of Ti fraction. (b) Variation of $T_c$ as a function of sheet-resistance. The solid line is a fit with Finkel'stein's equation.}
	\label{fig:xRTc}
\end{figure}

In Fig.~\ref{fig:xRTc}a, we plot $R_{s}$, measured at 50\,K where free electron parameters are defined, as a function of $x$. Clearly $R_{s}$ decreases with $x$. 

To apply Finkelstein's equation, in Fig.~\ref{fig:xRTc}b, we plot $T_c$ as a function of $R_{s}$. The solid line is a fit, taking  $T_{c0}$ and $\tau$ as fit parameters. We extract  $T_{c0} = 16.1$\,K and $\tau = 5.9 \times 10^{-16}$\,s from the fit. $\tau$, as extracted from the fit, is of the same order of magnitude as estimated from the free electron theory (see Table-\ref{tab1}). $T_{c0}$, on the other hand, is close to bulk $T_{c}$ reported on Nb$_{1-x}$Ti$_x$N samples, which typically range between 16 to 17\,K \cite{hazra2016superconducting,di1990niboium}. However, we would like to point out that each sample, in principle, can have different $T_{c0}$ and $\tau$. Thus, the extracted  $T_{c0}$ and $\tau$ only represent average values.   

We would like to point out that Finkelstein's model is valid for 2d systems. Our films' thickness (10\,nm) is roughly twice $\xi(0)$ (see supplementary information for $\xi (0)$ measurement). Thus, our samples are not exactly in the 2d limit, but very close to it.

In summary, we report on the superconducting properties of five disordered Nb$_{1-x}$Ti$_x$N thin films of thickness 10\,nm with different $x$.  We see that  the disorder of the films decrease with increase of $x$. Consequently, with the increase of $x$, the disorder-induced Coulomb interaction reduces, leading to an increase of $T_c$. Our analysis shows quantitative agreement with the Finkelstein's theory of disordered superconductivity.

Acknowledgments--- We acknowledge financial support from the French National Research Agency/grant ANR-14-CE26-0007--WASI, from the Grenoble Nanosciences Foundation/grant JoQOLaT and
from the European Research Council under the European Union’s Seventh Framework Programme (FP7/2007-2013)/ERC Grant agreement No 278203--WiQOJo.  


\bibliography{Bibliography}
 
\end{document}